\documentclass[twocolumn,showpacs,preprintnumbers,amsmath,amssymb,superscriptaddress]{revtex4}

\usepackage{graphicx}
\usepackage{dcolumn}
\usepackage{bm}

\begin{document}

\preprint{APS/123-QED}

\title{Single-shot measurement of the Josephson charge qubit}

\author{O. Astafiev}
\email{astf@frl.cl.nec.co.jp} \affiliation{The Institute of
Physical and Chemical Research (RIKEN), Wako, Saitama 351-0198,
Japan}
\author{Yu. A. Pashkin}
\altaffiliation[On leave from ]{Lebedev Physical Institute, Moscow
117924, Russia} \affiliation{The Institute of Physical and
Chemical Research (RIKEN), Wako, Saitama 351-0198, Japan}
\author{T. Yamamoto}
\affiliation{The Institute of Physical and Chemical Research
(RIKEN), Wako, Saitama 351-0198, Japan} \affiliation{NEC
Fundamental Research Laboratories, Tsukuba, Ibaraki 305-8501,
Japan}
\author{Y. Nakamura}
\affiliation{The Institute of Physical and Chemical Research
(RIKEN), Wako, Saitama 351-0198, Japan} \affiliation{NEC
Fundamental Research Laboratories, Tsukuba, Ibaraki 305-8501,
Japan}
\author{J. S. Tsai}
\affiliation{The Institute of Physical and Chemical Research
(RIKEN), Wako, Saitama 351-0198, Japan} \affiliation{NEC
Fundamental Research Laboratories, Tsukuba, Ibaraki 305-8501,
Japan}

\date{\today}

\begin{abstract}
We demonstrate single-shot readout of quantum states of the
Josephson charge qubit. The quantum bits are transformed into and
stored as classical bits (charge quanta) in a dynamic memory cell
- a superconducting island. The transformation of state
$|1\rangle$ (differing form state $|0\rangle$ by an extra Cooper
pair) is a result of a controllable quasiparticle tunneling to the
island. The charge is then detected by a conventional
single-electron transistor, electrostatically decoupled from the
qubit. We study relaxation dynamics in the system and obtain the
readout efficiency of 87\% and 93\% for $|1 \rangle$ and $|0
\rangle$ states, respectively.
\end{abstract}

\maketitle

It has been recently realized that Josephson junctions can be used
for building quantum bits (qubits) and integrated quantum computer
circuits controlled by external electrical signals
\cite{Makhlin,Averin}. After the first experiments on single
Josephson qubits \cite{Nakamura,Vion,Yu,Martinis,Chiorescu}, an
important step towards the integration has been made: coherent
control of two qubits and conditional gate operation have been
experimentally demonstrated for two electrostatically coupled
charge qubits \cite{Pashkin,Yamamoto}. However, in these
experiments, individual probabilities of each qubit averaged over
all states of the other qubit were measured \cite{Qstate}. To
directly measure multi-qubit states, one must be able to readout
each qubit after every single-shot coherent state manipulation.
The single-shot readout is of great importance, for instance, for
quantum state tomography, quantum state teleportation, quantum
cryptography \cite{Nielsen}. Without the single-shot readout,
algorithms that give non-unique solutions can not be utilized.

To readout single quantum states of the Josephson qubits (in
particular, flux qubits) through the phase degree of freedom, a
few circuits, measuring switching event from the supercurrent
state to the finite-voltage state were implemented
\cite{Vion,Yu,Martinis,Chiorescu}. In charge type of qubits, it is
straightforward to measure a charge quantum instead of the flux
quantum \cite{Shnirman}. For the single-shot charge readout, a
radio-frequency single-electron transistor \cite{Schoelkopf}
electrostatically coupled to the qubit was proposed as a detector
of the charge states \cite{Aassime,Johansson}. Although this
approach works in principle \cite{Duty,Lehnert}, the single-shot
readout has not yet been realized. In this work, we demonstrate an
operation and study mechanism of novel readout scheme that allows
to perform highly efficient single-shot measurements, with
suppressed back-action of the measurement circuit on the qubit.

\begin{figure}
\includegraphics{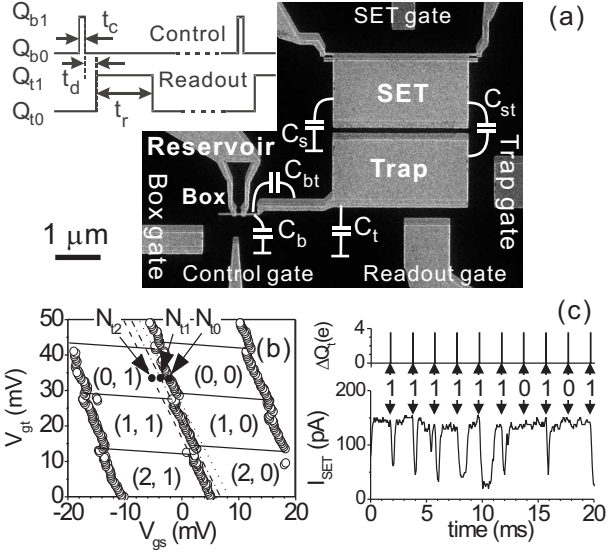}
\caption{\label{fig:AstafievFig1} (a) Scanning electron micrograph
of the device. The aluminum structure is deposited on top of a
thin Si$_{3}$N$_{4}$ insulating layer (0.4 $\mu$m) above a gold
ground plane. The device consists of a Cooper pair box, a
reservoir, a trap and a measurement SET. The dc ("box" and "trap")
and pulse (control and readout) gates control potentials of the
islands. Pulse operation is schematically represented in the
inset. (b) Stability diagram of the SET coupled to the trap. Open
circles mark positions of the SET quasiparticle current peaks on
$V_{gs} - V_{gt}$ plane (dc gate voltages of the SET and the
trap). Pairs of numbers $(N_{t}, N_{s})$ designate the trap - SET
ground state charge configuration in each cell bounded by the SET
peaks and solid lines. Dashed, dashed-dotted and dotted lines
indicate positions of the SET peaks for 0, 1 or 2 additional
electrons in the trap, respectively. (c) A typical time-trace of
the SET current (lower panel) together with the readout pulse
sequence (upper panel). Negative switches on the lower curve
correspond to the detected charge of the trap. Digits 1 and 0 mark
readout bits. }
\end{figure}

A scanning-electron micrograph of our circuit is shown in
Fig.~\ref{fig:AstafievFig1}(a) The device consists of a charge
qubit \cite{Nakamura} and a readout circuit. The qubit is a
Cooper-pair box (with its effective capacitance to the ground
$C_{b} \approx$ 600 aF) coupled to a reservoir through a Josephson
junction with the Josephson energy $E_{J} \approx 20 \mu$eV. The
reservoir is a big island with about 0.1 nF capacitance to the
ground plane and galvanically isolated from the external
environment. The qubit states are coherently controlled by a
non-adiabatic control pulse, yielding a superposed state of $|0
\rangle$ and $|1 \rangle$. The readout part includes an
electrometer which is a conventional low-frequency single-electron
transistor (SET) ($C_{s} \approx$ 1000 aF) and a charge trap
($C_{t} \approx$ 1000 aF) placed between the qubit and the SET.
The trap is connected to the box through a highly resistive tunnel
junction ($R_{t} \approx$ 100 M$\Omega$) and coupled to the SET
with a capacitance $C_{st} \approx$ 100 aF. The use of the trap
enables us to separate in time the coherent state manipulation and
readout processes and, in addition, the qubit becomes
electrostatically decoupled from the SET. The qubit relaxation
rate induced by the SET voltage noise is suppressed by a factor of
$(C_{bt} C_{st}/C_{t} C_{b})^{2} \approx 3\times10^{-5}$, where
$C_{bt} \approx$ 30 aF \cite{Relaxation}. The coupling strength
can be made even weaker, if the unwanted box-to-trap capacitance
$C_{bt}$ is further decreased.

The operation of the circuit can be described in the following
way. During the qubit manipulation, the trap is kept unbiased
prohibiting charge relaxation to the trap. Once the control pulse
is terminated, the readout pulse (see the inset of
Fig.~\ref{fig:AstafievFig1}(a)) is applied to the trap. The length
and the amplitude of the readout pulse are adjusted so that if
there is an extra Cooper pair in the box after termination of the
control pulse, it escapes to the trap through a quasiparticle
tunneling with a high probability. After the charge is trapped, it
remains in the trap for a long time (a reverse trap-to-box charge
relaxation is suppressed due to the superconducting energy gap
2$\Delta$) and is measured by a low-frequency SET.

The Hamiltonian of the two-level system of the qubit in the charge
basis $|0 \rangle$ and $|1 \rangle$ (without and with an extra
Cooper pair) is $H = U_{b}(0, Q_{b}) |0 \rangle \langle 0| +
U_{b}(2, Q_{b}) |1 \rangle \langle 1|$ - 1/2 $E_{J}$ ($|0
\rangle\langle 1|$ + $|1 \rangle \langle 0|$) (we define an
electrostatic energy of island $k$ as $U_{k}(N_{k}, Q_{k}) =
(N_{k}e - Q_{k})^{2}/2C_{k}$, where $k$ is either $b$ or $t$
indicating box or trap island, respectively, $N_{k}$ is an excess
electron number, and $Q_{k}$ is a gate induced charge in the
island. Starting at $Q_{b0} (Q_{b0} < Q_{b1}$), where $E
>> E_{J}$ ($\Delta E = U_{b}(2, Q_{b}) - U_{b}(0, Q_{b}))$ we let the system relax to
the ground state, which is nearly pure charge state $|0\rangle$.
Then we instantly change the eigenbasis for a time $t_{c}$ by
applying a rectangular control pulse, which brings the system to
$Q_{b1}$. If $Q_{b1}$ is a degeneracy point ($\Delta E = 0$), the
final state of the control pulse manipulation is $|0 \rangle \cos
\omega_{J} t_{c}/2 + |1\rangle \sin \omega_{J} t_{c}/2$
($\omega_{J} = E_{J}/\hbar$), therefore, after the pulse
termination, the state $|1\rangle$ is realized with a probability
of $\cos^{2} \omega_{J} t_{c}/2$.

Fig.~\ref{fig:AstafievFig1}(b) shows an experimentally measured
stability diagram: SET current peak positions as a function of
trap and box gate voltages of the SET and the trap. By setting the
box and trap gates to one of the points $N_{t0}$, $N_{t1}$ or
$N_{t2}$, we can detect if the trap has 0, 1 or 2 additional
electrons. In our measurements, the SET is usually adjusted to
$N_{t0}$. To readout the qubit, the trap is biased by the readout
pulse of typical length $t_{r}$ = 300 ns and amplitude $\Delta
Q_{t} = 3.5e$ ($\Delta Q_{t} = Q_{t1} - Q_{t0}$), applied to the
readout gate, letting an extra Cooper pair of the state $|1
\rangle$ escape to the trap through a quasiparticle tunneling and
switching off the SET current (the SET peak position is shifted to
the position of the dashed line).

The curve in the upper panel of Fig.~\ref{fig:AstafievFig1}(c)
indicates a readout pulse sequence. The curve on the lower panel
demonstrates a typical time-trace of the SET current. Negative
switches on the curve of the lower panel coming synchronously with
the readout pulses are counted as charge detection events. For the
studied device, the lifetime of the trapped charge is typically
about 300 $\mu$s, therefore, normally used repetition time of 2 ms
is sufficient for practically complete trap resetting. We count
the number of detected switches $m$, with the total number of
shots $n_{t}$.

\begin{figure}
\includegraphics{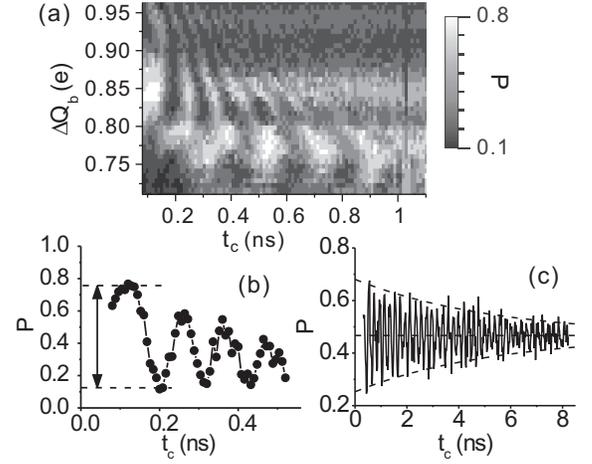}
\caption{\label{fig:AstafievFig2} (a) Coherent oscillations
measured by averaging over many events of the single-shot
measurements. (b) $P$ versus $t_{c}$ measured at $\Delta Q_{b} =
0.84e$ ($\equiv Q_{bA}$), where visibility is the highest. (c) $P$
versus $t_{c}$ measured at $Q_{b} = 0.75e$ ($\equiv Q_{bB}$),
where the oscillations are the longest lasting (degeneracy point).
Dashed envelops correspond to the exponential decay with the decay
time of 5.8 ns. }
\end{figure}

An experimentally obtained probability of the charge detection $P
= m/n_{t}$ (normally, $n_{t}$ = 327 is used per one experimental
data point) as a function of the control pulse length $t_{c}$ and
the amplitude $\Delta Q_{b}$ ($\Delta Q_{t} = Q_{t1} - Q_{t0}$) is
shown as a two-dimensional plot in Fig.~\ref{fig:AstafievFig2}(a).
We define the pulse with $\Delta Q_{b} = 0.84e$ ($\equiv Q_{bA}$)
and $t_{c}$ = 120 ps, when $P$ reaches maximum, as a $\pi$-pulse.
Fig.~\ref{fig:AstafievFig2}(b) demonstrates coherent oscillations
as a function of $t_{c}$ measured at $\Delta Q_{bA}$. As shown by
the vertical arrowed line, the visibility here reaches 0.64, while
the longest lasting oscillations shown in
Fig.~\ref{fig:AstafievFig2}(c) are found at $\Delta Q_{b} = 0.75e$
($\equiv Q_{bB}$), (the phase decoherence is expected to be the
weakest at the degeneracy point). We believe that $Q_{bA} \neq
Q_{bB}$ due to the control pulse distortion because of limited
frequency bandwidths of the transmission lines and the pulse
generator.

\begin{figure}
\includegraphics{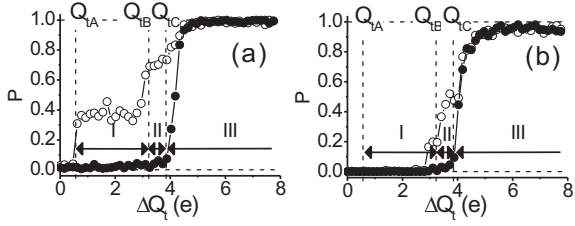}
\caption{\label{fig:AstafievFig3} $P$ versus readout pulse
amplitude $\Delta Q_{t}$ with and without $\pi$-pulse (open and
closed cycles, respectively). (a) $P$ measured at $N_{t0}$ (see
Fig.~\ref{fig:AstafievFig1}(b)). (b) $P$ measured at $N_{t2}$. One
may distinguish three different regions on the plots: (I) a finite
probability for the excited state detection is only at $N_{t0}$
position; (II) finite probabilities for the excited state
detection are at $N_{t0}$ and $N_{t2}$ positions; (III) switches
are detected even if the qubit is in the state $|0\rangle$.
Threshold amplitudes $Q_{tA}$, $Q_{tB}$, $Q_{tC}$ are derived for
the box-to-trap relaxation processes $(2, 0) \rightarrow (1, 1)$,
$(1, 1) \rightarrow (0, 2)$ and $(0, 0) \rightarrow (-1, 1)$,
respectively.}
\end{figure}

Curves with open and closed dots in
Figs.~\ref{fig:AstafievFig3}(a-b) represent $P$ as a function of
the readout pulse amplitude $\Delta Q_{t}$ with and without
control $\pi$-pulses, respectively. The probabilities measured at
$N_{t0}$ are shown in Fig.~\ref{fig:AstafievFig3}(a), while the
probabilities measured at $N_{t2}$ (where positive switches are
counted) are shown in Fig.~\ref{fig:AstafievFig3}(b)
.
One may divide the plots into three regions marked by I, II and
III differing from each other by the counting characteristics.
Based on the data from these plots, we suppose that when the qubit
is in the state $|1\rangle$, one quasiparticle tunnels from the
box to the trap in the process (2, 0) $\rightarrow$ (1, 1) within
the region I (($N_{b}, N_{t}$) represents the box-trap
quasiparticle configuration); quasiparticles tunnel to the trap in
the process (2, 0) $\rightarrow$ (1, 1) $\rightarrow$ (0, 2)
within the region II; a quasiparticle tunneling process becomes
possible even for the state $|0\rangle$ in the region III.

The quasiparticle tunneling is energetically feasible in the
process $(N_{b}, N_{t}) \rightarrow (N_{b} - 1, N_{t} + 1)$, when
the following condition is satisfied $U_{b}(N_{b}, Q_{b0}) +
U_{t}(N_{t}, Q_{t1})
> U_{b}(N_{b} - 1, Q_{b0}) + U_{t}(N_{t} + 1, Q_{t1}) + 2\Delta$ \cite{Parity} (we neglect
the interaction energy term which is as small as the inter-island
coupling strength). Substituting an explicit expression for the
energies, one may find the necessary trap readout pulse amplitude
for the quasiparticle escape  $\Delta Q_{t} > Q_{t}' + N_{t}e +
\eta (2 - N_{b}) e$, where $\eta = C_{t}/C_{b}$ and $Q_{t}'  = e/2
- \eta (3e/2 - Q_{b0}) + 2\Delta C_{t}/e - Q_{t0}$. We define
three threshold amplitudes of $\Delta Q_{t}$, at which the
following processes become possible: (2, 0) $\rightarrow$ (1, 1)
at $\Delta Q_{t} > Q_{tA}$, where $Q_{tA} = Q_{t}'$ ; (1, 1)
$\rightarrow$ (0, 2) at $\Delta Q_{t} > Q_{tB}$, where $\Delta
Q_{tB} = Q_{t}' + (1 + \eta)e$; (0, 0) $\rightarrow$ (-1, 1) at
$\Delta Q_{t} > Q_{tC}$, where $Q_{tC} = Q_{t}' + 2 \eta e$. The
threshold amplitudes calculated with $Q_{tA} = 0.6e$ (taken from
the first step of $P$ curve on Fig.~\ref{fig:AstafievFig3}(a)) and
$\eta$ = 1.67 are $Q_{tB} = 3.3e$ and $Q_{tC} = 3.9e$ and shown on
the Figs.~\ref{fig:AstafievFig3}(a-b) by the dashed vertical
lines. Note that the probability for the trap to have a charge at
$\Delta Q_{t} > 4.5e$ reaches 1 in Fig.~\ref{fig:AstafievFig3}(a).
This proves that once the charge is trapped it is detected with
100\% probability. For the highest efficiency, we set the
operation point of the SET to $N_{t0}$, where either one or two
trapped quasiparticles yield a negative switch of the SET current
and the pulse amplitude to $\Delta Q_{t} \approx 3.5e$, at which
two quasiparticles may escape to the trap.

\begin{figure}
\includegraphics{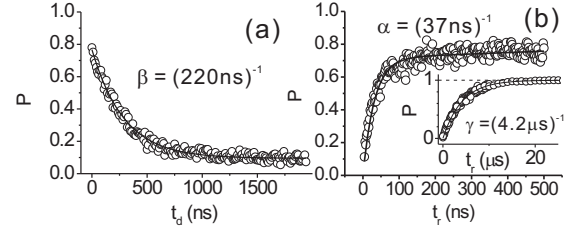}
\caption{\label{fig:AstafievFig4} (a) $P$ as a function of delay
between the control $\pi$-pulse and the readout pulse $t_{d}$. A
solid curve is a fitting by an exponent with a decay rate $\beta$
= (220 ns)$^{-1}$. (b) A probability of $|1\rangle$-states
detection created by the -pulse as a function of the readout pulse
length $t_{r}$. A solid curve is a result of fitting $m(t)$ using
Eq. (\ref{eq:Eq1}) normalized by $n_{t}$ with fitting parameters
$n_{0}^{*}$ and $\alpha$ ($n_{0}^{*}/n_{t} = 0.87$ and $\alpha$ =
(37 ns)$^{-1}$). The inset shows probability without $\pi$-pulses.
$P$ is fitted by 1 - $\exp(- \gamma t_{r})$ with $\gamma$ = (4.1
$\mu$s)$^{-1}$. }
\end{figure}

Figs.~\ref{fig:AstafievFig4}(a-b) demonstrate time relaxation
dynamics of the qubit states. Fig.~\ref{fig:AstafievFig4}(a) shows
a probability $P$ to find an extra charge in the trap, when time
delay $t_{d}$ is introduced between the control $\pi$-pulse and
the readout pulses. The exponential decay of $P$ may be explained
by tunneling to the reservoir (presumably via energetically
feasible Cooper pair tunneling (2, 0) $\rightarrow$ (0, 0))
because alternative quasiparticle relaxation to the trap through
the high resistive junction is blocked by 2$\Delta$ when the trap
is not biased. The relaxation rate to the reservoir found from the
fitting (solid curve) is $\beta$ = (220 ns)$^{-1}$.
Fig.~\ref{fig:AstafievFig4}(b) shows relaxation dynamics of the
state $|1\rangle$ as a function of the readout pulse length
$t_{r}$ ($t_{d} \approx 0$). This relaxation is mainly determined
by quasiparticle decay to the trap with a rate  $\alpha$ ($\alpha
\gg \beta$). Additionally, the inset of
Fig.~\ref{fig:AstafievFig4}(b) shows dynamics of $|0\rangle$-state
relaxation ("dark" switches) to the trap. These "dark" switches
can be presumably described by the process (0, 0) $\rightarrow$
(-2, 2), with a weak relaxation rate $\mu$ = (4100 ns)$^{-1}$
derived from fitting the data by 1 - $\exp(- \gamma t_{r})$ (solid
curve).

Let us consider the relaxation dynamics in more details. The
number of excited states, $n^{*}$, decreases within the time
interval $[t, t + dt]$ as $dn^{*}(t) = - \alpha  n^{*}(t) dt -
\beta n^{*}(t) dt$ \cite{Nstar}. The number of states in (0, 0)
configuration, $n(t)$, changes, in turn, as $dn(t) = - \gamma n(t)
dt + \beta n^{*}(t) dt$. We may also write an expression for the
number of events in which the trap is found to be charged: $dm(t)
= n^{*}(t) dt + n(t) dt$. Solving these equations with the initial
conditions $n^{*}(0) = n_{0}^{*}$, $n(0) = n_{t} - n_{0}^{*}$ and
$m(0) = 0$ we find
\begin{equation}
m(t) = n_{t}[1- e^{-\gamma t}]
 +  \frac{n_{0}^{*}(\alpha - \gamma)}{\alpha + \beta - \gamma}
[e^{-\gamma t}- e^{-(\alpha + \beta)t}]. \label{eq:Eq1}
\end{equation}
We fit the data of Fig.~\ref{fig:AstafievFig4}(b) by a curve of $P
= m(t)/n_{t}$ with $m(t)$ taken from Eq. (\ref{eq:Eq1}) with two
fitting parameters $\alpha$ and $n_{0}^{*}$. The fitting gives
$\alpha$ = (37 ns)$^{-1}$ \cite{Resistance} and $n_{0}^{*}/n_{t}$
= 0.84, implying that the efficiency of $|0\rangle$-to-$|1\rangle$
conversion by the control $\pi$-pulse is 84\%.

If our readout pulse length $t = t_{r}$ satisfies the conditions
$\gamma \ll t_{r}^{-1} \ll \alpha + \beta$, then Eq.
(\ref{eq:Eq1}) can be simplified to
\begin{equation}
m(t) \approx n_{0}^{*} \frac{\alpha}{\alpha + \beta} +\gamma
t_{r}[n_{t} - n_{0}^{*}\frac{\alpha}{\alpha + \beta}].
\label{eq:Eq2}
\end{equation}
Using Eq. (\ref{eq:Eq2}), one may estimate an efficiency of the
single-shot readout. We introduce a probability $P_{y}(x)$ of
finding the trap charged ($y$ = 1) or uncharged ($y$ = 0), when
the qubit is in $|x\rangle$-state ($x$ is either 0 or 1).
According to the definition of $P_{y}(x)$, $P_{0}(0) + P_{1}(0) =
1$ and $P_{0}(1) + P_{1}(1) = 1$. The total number of detected
events expressed in terms of these probabilities may be written as
$m = n_{0}^{*} P_{1}(1) + (n_{t} - n_{0}^{*}) P_{1}(0)$. Comparing
the latter expression with Eq. (\ref{eq:Eq2}) we find
\begin{subequations}
\label{eq:Eq3}
\begin{eqnarray}
P_{1}(1) = \frac{\alpha + \beta \gamma t_{r}}{\alpha +
\beta}\label{eq:1}
\\
P_{1}(0) = \gamma t_{r}.\label{eq:2}
\end{eqnarray}
\end{subequations}

Confirming that our readout pulse length $t_{r}$ = 300 ns fulfills
the necessary condition for Eqs. (\ref{eq:Eq3}), $\gamma \ll
t_{r}^{-1} \ll \alpha + \beta$, we directly find from Eqs.
(\ref{eq:Eq3}) that the probability of detection of the state
$|1\rangle$ is $P_{1}(1)$ = 0.87 and the state $|0\rangle$ is
$P_{0}(0)$ = 0.93 ($P_{1}(0) = 0.07$). The readout efficiency can
be further improved by optimizing the relaxation rates. The
derived probabilities are consistent with the mean probability of
the oscillations at the degeneracy point, $\langle P \rangle =
[\langle n_{0}^{*}\rangle P_{1}(1) + (n_{t} - \langle
n_{0}^{*}\rangle) P_{1}(0)]/n_{t} = 0.47$ (horizontal line in
Fig.~\ref{fig:AstafievFig4}(c)), where $\langle
n_{0}^{*}/n_{t}\rangle = 0.5$.

Finally, we would like to briefly discuss new possibilities and
potential applications, which this scheme opens. The trap island
works as a dynamic memory cell, which stores the multi-qubit state
in a classical bit. Therefore, the data can be further processed
in a classical way. For example, implementing sequential readout
of many cells connected in series, similarly to that of
charge-coupled devices will allow to reduce a number of output
circuit channels.

One of straightforward applications of the described device,
staying outside quantum computation direction, is a tunable
single-photon detector in a centimeter wavelength range. A weak
microwave radiation may resonantly excite the box near the
degeneracy point from the ground to the excited state with a
finite probability. The excited state may then be converted into
the charge state $|1\rangle$ by an adiabatic sweep of the control
gate voltages (a relatively slow gate voltage change forces the
system to move along eigenenergy bands) and detected by the
single-shot measurement circuit.

As for the nearest plans, it is straightforward to use the
single-shot readout to for measuring double qubit system
\cite{Pashkin,Yamamoto}. This will allow us to demonstrate
entangled states and controlled-NOT gate operation.

We thank S. Lloyd for valuable discussion.


\begin{thebibliography}{21}
\bibitem[{Qst()}]{Qstate}
\bibinfo{note}{Quasiparticle currents induced in the measurement circuits of
  the qubits are proportional to individual probabilities of finding the qubits
  in the state $|1\rangle$ (with one extra Cooper pair). That is, if the
  wave-function in the charge basis of the double qubit basis of
  $|i_{1}i_{2}\rangle$ is $\sum_{i_{1}i_{2}} C_{i_{1}i_{2}} |i_{1}i_{2}\rangle$
  ($i_{k}$ are 1 or 0 depending on whether the qubit $k$ has an extra Cooper
  pair or not), the current induced, for example, in the qubit 1 measurement
  circuit is proportional to $\sum_{i_{2}} \vert C_{i_{1}i_{2}}\vert ^{2}$.}
\bibitem[{Rel()}]{Relaxation}
\bibinfo{note}{We estimate the qubit energy relaxation rate during coherent
  state manipulation induced by our measurement SET to be lower than (0.5
  $\mu$s)$^{-1}$.}
\bibitem[{Par()}]{Parity}
\bibinfo{note}{In our case of the high resistive junction and relatively short
  readout pulses, an additional energy 2$\Delta$ is needed for quasiparticle
  tunneling independently of parity because of low tunneling rate of an
  unpaired quasiparticle.}
\bibitem[{Nst()}]{Nstar}
\bibinfo{note}{We simplify this problem avoiding consideration of the detailed
  relaxation mechanism and assume that $n^{*}$ includes any ($N_{b}$, 0) charge
  configurations with $N_{b} \neq 0$. Most importantly, the experimentally
  measured decay of $n^{*}$ in Fig.~\ref{fig:AstafievFig4}(a) is perfectly
  exponential.}
\bibitem[{Res()}]{Resistance}
\bibinfo{note}{The tunnel junction resistance is estimated as $R_{t} \approx
  2\Delta/\alpha e^{2}$ = 100 $\Omega$M .}
\end{thebibliography}

\end{document}